\documentclass[ ]{revtex4}
\usepackage{hyperref}
\usepackage{amsmath}

\usepackage{amsfonts}

\begin{document}
\title{Finite Field-dependent Symmetry in Thirring Model} 

 \author { Sudhaker Upadhyay }
 \email{ sudhakerupadhyay@gmail.com}
\affiliation {  Center for Theoretical Studies,
Indian Institute of Technology  Kharagpur, Kharagpur-721 302, India}
 
 \author { Prince A.  Ganai }
 \email{ princeganai@nitsri.net}
\affiliation { Department of Physics, National Institute of Technology,  Srinagar, Kashmir-190006, India } 
\begin{abstract}
 In this paper, we consider a $D$-dimensional massive Thirring model with ($2<D<4$). We derive an extended
BRST symmetry of the theory with finite field-dependent parameter. Further
we compute the Jacobian of functional measure under such an extended transformation.
Remarkably, we find that such Jacobian extends the BRST exact part of the action which leads to a mapping between different gauges. We illustrate this with the help of Lorentz and $R_\xi$ gauges.
We also discuss the results in  Batalin-Vilkovisky framework.
\end{abstract}

\maketitle
\section{Introduction}
From earlier studies in quantum field theories (QFT), it is found that ground states showing sensitivity to
the number of light fermion flavors  are very important. One of the
major example of such theories is the
 Thirring model, a quantum field theory of fermions interacting via a  conserved vector current term, 
 described in three-dimensional space-time.  
   Thirring model  has been studied as
a candidate for the scenario of the fermion dynamical mass generation \cite{a,b,c}.
 It is important because the fermion dynamical mass generation is the central issue of the
dynamical electroweak symmetry breaking such as the technicolor \cite{wei0}  and the top
quark condensate \cite{vo}.  The role of four fermion
interaction in the context of walking technicolor \cite{3} and  strong ETC technicolor \cite{4}
has also been investigated. In particular, the scalar/pseudoscalar-type four fermion interactions with 
the gauge interaction have played
a very important role in  $D= 4$ dimensions  as a renormalizable model \cite{u}.  This model with
gauge interaction
is known as gauged Nambu-Jona-Lasinio  (NJL) model \cite{x,y,z}. It has
 been observed  that the phase structure of such a gauged NJL model in  $D= 4$ dimensions is quite similar to that of the $D=3$   dimensional scalar/pseudoscalar-type   four fermion theory  
without gauge interactions  \cite{8},     called as Gross-Neveu model \cite{ga}.
The gauged Thirring model,  a natural gauge invariant generalization
of the Thirring model, has been studied in \cite{kond}, where it is shown that, in the strong gauge-coupling limit, the gauged 
Thirring model reduces to the  
proposed reformulation of the Thirring model  \cite{b} as a gauge theory.
 
To quantize the gauge theories,  
  the BRST formulation  \cite{brst, tyu,wei,sudu}  is a  powerful method,   which guarantees the renormalizability and  unitarity of  gauge  theories. 
  In \cite{c} certain aspects of BRST quantization for Thirring model in ($2<D<4$) dimensions are discussed. 
  This model is described in Lorentz gauge and $R_\xi$ gauges there.
  For instance, it is well-known that,  in $R_\xi$ gauge,  the   Stueckelberg (also 
  called  Batalin-Fradkin) field    $\theta$   is completely 
decoupled to the massive vector boson independently of $\xi$. This would lead to
simplicity in performing a numerical computations.  However, 
for Lorentz gauge, the Stueckelberg field    $\theta$ is  decoupled to  the massive vector boson 
only for $\xi=0$ (which refers Landau gauge). 
  These two gauges have their own advantages. A mapping between these two gauges 
 in the perspective of  the Thirring model would be remarkable because if one gets a complicated expression for the calculations in one gauge, then this mapping would be very helpful. In this context, we try to achieve this goal  with the help of generalization
   of BRST quantization and with the Batalin-Vilkovisky (BV) description.  
   
  The key idea of  generalization (extension) of BRST symmetry 
is to make the parameter of transformation finite and   field-dependent in certain way, known as finite field-dependent BRST (FFBRST) transformation \cite{sdj}.
The generalization, in this way,  has found 
enormous applications in wide area of gauge theories \cite{sdj1,susk} as well as in gravity theory \cite{upad}. For example, the celebrated Gribov issue \cite{gri, zwan,zwan1} in Yang-Mills theory
has been addressed in the framework of FFBRST formulation (for details see refs. therein \cite{sb}). 
The FFBRST transformations have been emphasized in higher-form gauge theories, 
an important ingredient of string theories \cite{smm}.  
Further, for the  superconformal Chern-Simmons- matter theories \cite{bl,g,abjm}, the aspects of
the generalized BRST symmetry have also reported in \cite{ fs,sudd,fsm}. The validity of 
such generalization has been established at quantum level also  \cite{ssb,sud001} with the help of
the BV formulation \cite{ht}. Recently, the FFBRST formulation has been studied in 
(topological) lattice sigma models \cite{rs}. A slightly different field-dependent BRST
formulation  has also been made, in Yang-Mills theories also  \cite{lav},
which involves a linear   dependence on  the corresponding Grassmann-odd parameter,
 naturally, without having recourse to any  quadratic dependence. Since Ref. \cite{lav} does not
 deal with the case of BRST-antiBRST symmetry,  and so any non-trivial quadratic dependence
 on the transformation parameters cannot occur.   Moshin and Reshetnyak, in Ref. \cite{ale},   incorporates, systematically, the case  of BRST-antiBRST symmetry in Yang-Mills theories
 within the context of finite transformations, which deals
 with the case of a quadratic   dependence  on the corresponding parameters for two
 reasons: 1) finite BRST-antiBRST symmetry  does admit a non-trivial quadratic dependence
 on two different Grassmann-odd parameters,  2) this dependence actually turns out
 to be necessary for a systematic treatment  of finite BRST-antiBRST transformations.
 Further,   the concept of finite BRST-antiBRST
 symmetry to the case of general gauge theories has been extended in Refs. \cite{ale1, mos1},
 whereas Ref. \cite{mos} by the same authors
 generalizes the corresponding parameters
 to the case of arbitrary  Grassmann-odd
 field-dependent parameters, as compared
 to the so-called ``potential" form of parameters
 used in the previous articles \cite{ale,ale1,mos1}.

 The field theoretic models, with  fermion
 interactions of current-current type, are not renormalizable in $D=4$ as the coupling constant takes the dimension of
 mass inverse square. Nevertheless, it has been established
that a class of $D$ ($2<D<4$) dimensional four-fermion model   is renormalizable in the different expansion scheme \cite{k}.
In this paper, we consider a renormalizable $D$ ($2<D<4$) dimensional gauge non-symmetric Thirring model  to
discuss the various gauge connection. After the introduction of the auxiliary field,  the theory still remains  
gauge non-invariant. Of course, the Thirring model can be  rewritten into the massive vector theory with which
the fermion couples minimally.
First of all, we
discuss  the gauge invariant version of the model, which can be quantized correctly only after
breaking the local gauge invariance. This is achieved by fixing the  gauges  specifically.
For the present case, the Lorentz and $R_\xi$ gauges are considered.
We write corresponding gauge fixed Faddeev-Popov action. The resulting action, by summing
the classical action to the gauge-fixed action, remains invariant
under the BRST symmetry. Further, we generalize the BRST symmetry, by forming
the transformation parameter finite and field dependent. The generalized BRST symmetry
leaves the Faddeev-Popov action invariant. The only difference, with the usual BRST 
symmetry, is  the functional measure, which is not covariant under 
the generalized BRST transformation. So, we compute the Jacobian for functional measure
and found that it depends, explicitly,  on the infinitesimal field-dependent 
parameter. Then, the different value of the field-dependent  parameter will lead to
different contribution in the generating functional. Here, we show that, 
for a particular value of such parameter, Jacobian switches the generating 
functional from one gauge to another gauge. We illustrate this result for a particular set 
of gauges, namely, the Lorentz gauge and $R_\xi$ gauge. Further, we establish the result at
quantum level, by mapping the solutions of quantum master equation in BV framework.

The paper is organized as follows.
In section II, we discuss the BRST quantization of 
Thirring model, with an arbitrary as well as specific gauge choices in $D$ ($2<D<4$) dimensions. 
Then, we derive methodology for extended BRST symmetry for Thirring model, where 
Jacobian for path integral measure is computed explicitly. The connections of various gauges
through this extended BRST formulation are described in section IV. 
In section V, we establish the mapping of different gauges in BV formulation.
The last section summarizes  the present investigations with future motivations. 

\section{Thirring model: BRST symmetry}
\par
In this section, we analyse the Thirring model
$D$ ($2<D<4$) dimensions.
The Thirring model is given by the Lagrangian density  
\begin{eqnarray}
 {\cal L}_{Th}
 = \bar \psi^a i \gamma^\mu \partial_\mu \psi^a
 - m \bar \psi^a \psi^a
 - {G \over 2N}(\bar \psi^a \gamma^\mu \psi^a)^2.
 \label{Thirring}
\end{eqnarray}
Here, $\psi^a$ refers  a Dirac fermion with flavor index  
$a$ which runs from 1 to $N$. The gamma matrices $\gamma_\mu
(\mu=0,1,2,...,D-1)$ satisfy  the Clifford algebra, following
$\{ \gamma_\mu, \gamma_\nu \} = 2 g_{\mu\nu} {\bf 1}$.

The Lagrangian can, further, be rewritten in terms of  an auxiliary vector field
$A_\mu$ as  
\begin{eqnarray}
 {\cal L}_{Th'} = \bar \psi^a  i \gamma^\mu (\partial_\mu
 - {i \over \sqrt{N}} A_\mu) \psi^a
 - m \bar \psi^a \psi^a
 + {1 \over 2G} A_\mu^2,
 \label{th'}
\end{eqnarray}
which coincides with (\ref{Thirring}) when we perform equation of motion of $A_\mu$. Here, we note that
 the field   $A_\mu$ is just a vector field which represents the fermionic
current and does not transform as a gauge field. This Lagrangian does not have any gauge symmetry.
Besides the lacks the kinetic term for the Yang-Mills field, the theory given by above Lagrangian 
is identical with the massive Yang-Mills
theory.

The gauge invariant version of the Lagrangian is obtained, by introducing the Stueckelberg field $\theta$ which identified with the BF field as shown in \cite{lo},  as 
\begin{eqnarray}
 {\cal L}_{Th''}
 = \bar \psi^a i \gamma^\mu (\partial_\mu
 - {i \over \sqrt{N}} A_\mu) \psi^a
 - m \bar \psi^a \psi^a
 + {1 \over 2G}(A_\mu - \sqrt{N}\partial_\mu \theta)^2.
\label{th''}
\end{eqnarray}
The original Thirring model can be assumed as the  gauge fixed version of this gauge invariant 
Lagrangian \cite{nv}, 
which possesses the following $U(1)$ gauge symmetry:
\begin{eqnarray}
&&\delta \psi_a =( e^{i\phi}-1) \psi_a,\\
&& \delta A_\mu  = \sqrt{N} \partial_\mu \phi,\\
&& \delta \theta  =   \phi.
\end{eqnarray}
Here, $\phi$ denotes a fictitious Nambu-Goldstone boson field,  which  has to be absorbed
into the longitudinal component of $A_\mu$.

To quantize covariantly a gauge invariant theory, we need to break the local gauge invariance. 
It removes the fictitious degrees of freedom associated with the theory.
This can be achieved by restricting the gauge fields by a general gauge condition,
$ \Omega=\mathcal{F}[A,\theta]=0$.
This can be incorporated at the level of Lagrangian by adding following linearized 
gauge-fixing and ghost terms to 
the classical action,
\begin{eqnarray}
 {\cal L}_{GF+FP}
 = B\mathcal{F}[A,\theta]+\frac{\xi}{2}B^2+i\bar C \left( \frac{\delta \mathcal{F}[A,\theta]}{\delta A_\mu}\partial_\mu + \frac{1}{\sqrt{N}} \frac{\delta \mathcal{F}[A,\theta]}{\delta \theta }\right)C,
\end{eqnarray}
where $B$ is a Nakanishi-Lautrup type multiplier field and $\xi$ is an arbitrary gauge parameter.
\subsection{Lorentz gauge}
Now for a particular choice, so-called Lorentz gauge,   $ \mathcal{F}[A,\theta]=\partial^\mu A_\mu$,
the above gauge fixed Lagrangian  is alleviated  to,
\begin{eqnarray}
{\cal L}^L_{GF+FP}
 = B\partial^\mu A_\mu +\frac{\xi}{2}B^2+i \bar C \left[\partial_\mu\partial^\mu\right]C.
\end{eqnarray}
The effective action  for Thirring model in Lorentz gauge is given by
\begin{eqnarray}
 {\cal L}_{Th''}+{\cal L}^L_{GF+GH}&=&\bar{\psi}^a i \gamma^\mu(\partial_\mu-\frac{i}{\sqrt{N}}A_\mu)\psi^a-  m\bar{\psi}^a\psi^a +\frac{1}{2G}(A_\mu-\sqrt{N}\partial_\mu \theta)^2+B\partial^\mu A_\mu\nonumber\\
 &&+  \frac{\xi}{2}B^2+i\bar{C}\partial_\mu\partial^\mu C.
\end{eqnarray}
Here, we observe that the Stueckelberg field (BF)   $\theta$ is  coupled to  field
except for $\xi=0$ (Landau gauge).

This action is invariant under following BRST transformation:
\begin{eqnarray}
&& \delta_b A_\mu(x) =  -\partial_\mu C\eta,
\nonumber\\
&& \delta_b B = 0,
\ \ \ 
 \delta_b C= 0,
\nonumber\\
&& \delta_b \bar C =  i B\eta,
\ \ \delta_b \theta = -{1 \over \sqrt{N}} C\eta ,
\nonumber\\
&& \delta_b \psi^j(x)=  {i \over \sqrt{N}}  C \psi^j \eta,\label{brs}
\end{eqnarray}
where $\eta$ is an infinitesimal Grassmann parameter.
\subsection{$R_\xi$ guage}

For another important choice of gauge, $ \mathcal{F}[A,\theta]=\partial_\mu A^\mu+\sqrt{N}\frac{\xi}{G}\theta $,  so-called $R_\xi$ gauge, the gauge fixed Lagrangian is given by
\begin{eqnarray}
{\cal L}^R_{GF+FP}
 = B (\partial_\mu A^\mu+\sqrt{N}\frac{\xi}{G}\theta)+\frac{\xi}{2}B^2+i \bar C 
 \left[\partial_\mu\partial^\mu+\frac{\xi}{G}\right]C. 
\end{eqnarray}

Thus, the Faddeev-Popov  effective action for Thirring model in  $R_\xi$ gauge is given by
\begin{eqnarray}
 {\cal L}_{Th''}+{\cal L}^R_{GF+GH}&=&\bar{\psi}^a i \gamma^\mu(\partial_\mu-\frac{i}{\sqrt{N}}A_\mu)\psi^a -  m\bar{\psi}^a\psi^a  +\frac{1}{2G}(A_\mu-\sqrt{N}\partial_\mu \theta)^2  
+B (\partial_\mu A^\mu+\sqrt{N}\frac{\xi}{G}\theta)\nonumber\\
 && +\frac{\xi}{2}B^2+i \bar C \left[\partial_\mu\partial^\mu+\frac{\xi}{G}\right]C.  
\end{eqnarray}
 This, further, reduces to,
\begin{eqnarray}
 {\cal L}_{Th''}+{\cal L}^R_{GF+GH}&=&\bar{\psi}^a i \gamma^\mu(\partial_\mu-\frac{i}{\sqrt{N}}A_\mu)\psi^a -  m\bar{\psi}^a\psi^a -\frac{1}{2\xi}(\partial_\mu A^\mu)^2+\frac{1}{2G}A_\mu^2 \nonumber\\
 && -\frac{1}{2}\frac{N\xi}{G^2}\theta^2+\frac{N}{2G}(\partial_\mu \theta)^2+
 i\bar C \left[\partial_\mu\partial^\mu+\frac{\xi}{G}\right]C.
\end{eqnarray}
Here, we see that the   Stueckelberg   field  $\theta$   is completely
decoupled independently of $\xi$.
The effective action, in $R_\xi$ gauge, is also invariant under the same set of BRST transformation 
(\ref{brs}).
\section{Extended BRST transformation}
  In this section, we derive the extended BRST formulation, by making the parameter finite and field dependent, known as FFBRST transformation, at a general ground. 
 We, first, define BRST transformation of  a generic field $\phi(x)$ as follows:
  \begin{eqnarray}
  \phi(x)\longrightarrow \phi'(x)=\phi (x)+s_b \phi(x)\ \eta,\label{a}
  \end{eqnarray}
  where $s_b\phi$ refers to Slavnov variation and $\eta$ is 
  an infinitesimal anti-commuting global parameter.
  It is well known that, under such transformation, the  path integral measure as well as effective action remain invariant \cite{wei}.
  
Now, we interpolate a continuous
  parameter  ($\kappa;  0\leq \kappa\leq1$) through the fields $\phi(x)$   such that
 the $\phi(x, \kappa =0) =\phi (x)$  is the original field and, however,  $\phi(x, \kappa =1) =\phi' (x)=\phi (x)+ s_b\phi(x) \Theta [\phi]$ is the FFBRST transformed field, where $\Theta [\phi]$
 is finite field-dependent parameter. Such FFBRST transformation is justified by  the following
  infinitesimal field-dependent BRST transformation: \cite{sdj}
  \begin{eqnarray}
  \frac{d\phi(x, \kappa)}{d\kappa} =s_b \phi(x,\kappa) \Theta'[\phi (x,\kappa)].\label{b}
  \end{eqnarray}
    Now, integrating the above equation w. r. to  $\kappa$ from 0 to 1,   we get 
the FFBRST transformation,
\begin{eqnarray}
 \delta_b \phi (x) = \phi' (x) -\phi (x) = s_b\phi (x)\Theta [\phi(x)],\label{c}
  \end{eqnarray}
  where   the finite field-dependent parameter  is given by
\begin{equation}
\Theta [\phi] = \Theta ^\prime [\phi] \frac{ \exp f[\phi]
-1}{f[\phi]},
\label{80}
\end{equation}
 and $f[\phi]$ is given 
 by 
 \begin{eqnarray}
 f[\phi]= \sum_i \int d^4x \frac{ \delta \Theta ^\prime [\phi]}{\delta
\phi_i(x)} s_b \phi_i(x).
 \end{eqnarray} 
The FFBRST transformations, with field-dependent parameter, are
 also symmetry of the effective action but the cost we pay is that, they are no more nilpotent.
 Contrary to usual BRST symmetry,   they do not leave the functional measure invariant. Eventually,  the path integral measure under such transformation
changes non-trivially,  leading to a  local Jacobian
in the functional integration.  
So our goal here is  to compute the explicit Jacobian of the functional measure, under such a transformations. 
\subsection{Jacobian for field-dependent BRST transformation}
To compute the Jacobian for path integral measure, 
under the FFBRST transformation with an arbitrary  parameter $\Theta$,
we first define   the  generating functional 
for the Thirring model described by an effective action $S^{FP}_{Th}[\phi]$ as follows,
\begin{eqnarray}
 Z[0]=\int {\cal D}\phi\ e^{iS^{FP}_{Th}[\phi]},\label{zen}
 \end{eqnarray}
where ${\cal D}\phi$ refers the full functional measure.
Furthermore, we write the  functional measure under
the action of FFBRST transformation as follows  \cite{sdj}  
 \begin{eqnarray}
 {\cal D}\phi  = J(\kappa) {\cal D}\phi (\kappa) = J(\kappa +d\kappa) {\cal D}\phi (\kappa +d\kappa).
 \end{eqnarray}
Since,  this transformation is infinitesimal, so the transformation from $\phi(\kappa)$ to $\phi(\kappa+d\kappa)$ can, further,  be written  as  \cite{sdj}  
  \begin{eqnarray}
 \frac{J(\kappa)}{J(\kappa +d\kappa) }  = \sum_\phi\pm \frac{{\delta}\phi (\kappa +d\kappa)}{{\delta}\phi (\kappa)},
 \end{eqnarray}
 where the $+$ sign  is used for bosonic fields and $-$ is used for the fermionic fields.
 Utilizing  Taylor expansion around $\kappa$ in the above expression, we get the following 
 identification \cite{sdj}:
 \begin{eqnarray}
  \frac{1}{J}\frac{dJ}{d\kappa} d\kappa =- d\kappa\int d^Dx \sum_\phi\pm s_b\phi(x,\kappa) \frac{\delta\Theta'[\phi(x,\kappa)]}{\delta\phi(x,\kappa)}.
 \end{eqnarray}
This reduces to
 \begin{eqnarray}
 \frac{d\ln J[\phi]}{d\kappa} =-\int d^Dx \sum_\phi\pm s_b\phi(x,\kappa) \frac{\delta\Theta'[\phi(x,\kappa)]}{\delta\phi(x,\kappa)}.\label{cu}
 \end{eqnarray}
  To attain the expression for the finite Jacobian, from the (above)  infinitesimal one, we integrate  it over $\kappa$ with limits from $0$ to $1$. This leads to a logarithmic series,   
  \begin{eqnarray}
  \ln J [\phi]  &=&-\int_0^1 d\kappa\int d^Dx \sum_\phi\pm s_b\phi(x,\kappa) \frac{\delta\Theta'[\phi(x,\kappa)]}{\delta\phi(x,\kappa)},\nonumber\\
  &=&- \left(\int d^Dx \sum_\phi\pm s_b\phi(x) \frac{\delta\Theta'[\phi(x)]}{\delta\phi(x)}\right).
 \end{eqnarray}
Now, exponentiating the above relation leads to the 
expression for Jacobian for functional measure, 
under FFBRST transformation with an arbitrary 
field dependent parameter $\Theta'$, as follows
   \begin{eqnarray}
  J[\phi]   = {  \exp\left(-\int d^Dx \sum_\phi\pm s_b\phi(x) \frac{\delta\Theta'[\phi(x)]}{\delta\phi(x)}\right)}.\label{J}
 \end{eqnarray}
 Here, we see that  the Jacobian (\ref{J})  extends the Faddeev-Popov action (within functional integral) 
of the theory,  given in (\ref{zen}), as following:
 \begin{eqnarray}
 \int {\cal D}\phi'\ e^{iS^{FP}_{Th}[\phi']}=\int J[\phi] {\cal D}\phi \ e^{iS^{FP}_{Th}[\phi ]} =\int {\cal D}\phi \ e^{i\left(S^{FP}_{Th}[\phi ]+i\int d^Dx \left(\sum_\phi\pm s_b\phi 
 \frac{\delta\Theta'}{\delta\phi }\right)\right)}.
 \end{eqnarray}
 We will notice that the Jacobian amounts precise change in the BRST exact part of the 
 action, so, the dynamics of the theory does not change as the BRST exact part of a BRST 
 invariant function  does not alter the
 dynamics of the theory at cohomological level.
 
\section{Connection of Lorentz gauge to $R_\xi$ gauge} 
 In this section, we illustrate the results of section III with an specific example.
 Following the methodology discussed above, we first construct the FFBRST transformation for Thirring model,
\begin{eqnarray}
&& \delta_b A_\mu(x) =  -\partial_\mu C\Theta[\phi],
\nonumber\\
&& \delta_b B = 0,\ \  
 \delta_b C = 0,
\nonumber\\
&& \delta_b \bar C =  i B\Theta[\phi],
\nonumber\\
&& \delta_b \theta = -{1 \over \sqrt{N}} C \Theta[\phi],
\nonumber\\
&& \delta_b \psi^j(x) =  {i \over \sqrt{N}}  C \psi^j \Theta[\phi],
\end{eqnarray}
 where $\Theta[\phi]$ is an arbitrary finite field-dependent parameter
 satisfying $\Theta^2=0$.
 Now, we construct an specific   $\Theta$, described in terms of  $\Theta'$,
 to see the effect of FFBRST transformation in Thirring model. This is
 given by 
 \begin{eqnarray}
 \Theta'[\phi]= \int d^Dx \left[ \bar{C}\sqrt{N}\frac{\xi}{G}\theta\right]. 
 \end{eqnarray}
 For this choice of parameter, we calculate the Jacobian for functional measure 
   \begin{eqnarray}
  J[\phi]   = {  \exp\left(i\int d^Dx \left[ B\sqrt{N}\frac{\xi}{G}\theta +i\bar{C}\frac{\xi}{G}C\right] \right)},
 \end{eqnarray}
 where (\ref{J}) is utilized. 
 
 Here, we observe that the  Jacobian contributes to 
 the unphysical (BRST exact) part of the action.
 This Jacobian modifies the expression of generating functional 
 in Lorentz gauge  as follows
 \begin{eqnarray}
 \int {\cal D}\phi'\ e^{i \int d^Dx \left({\cal L}_{Th''}+{\cal L}^L_{GF+GH}\right) [\phi']} \stackrel{FFBRST}{---\longrightarrow}\int {\cal D}\phi \ e^{i \int d^Dx \left({\cal L}_{Th''}+{\cal L}^R_{GF+GH}\right) [\phi ]},
 \end{eqnarray}
 where the final expression is nothing but the generating functional in $R_\xi$ gauge.
 Such modification does not alter the theory because the extra pieces,
 due to Jacobian, attribute to the BRST exact part of the action.
 Though we have shown the connection of two specific gauges, this results are valid for any 
 arbitrary pair of gauges.
 Suppose, we choose a parameter $\Theta'[A, \theta, \bar C]= \int d^Dx \left[ \bar{C}\left(\mathcal{F}_1[A,\theta]-\mathcal{F}_2[A,\theta]\right)\right]$, where $\mathcal{F}_1$ and $\mathcal{F}_2$ are 
 two arbitrary gauges,
 then, the Jacobian will map the generating functional corresponding to these two gauges.
 Thus, we see that the two well studied gauges of Thirring model 
 are related to each other. This is shown with the helps of extended BRST transformation,
 with a particular parameter of transformation.
 \section{BV formulation and FFBRST symmetry} 
In  the BV formulation, the generating functional   of  Thirring model   (in Lorentz gauge),  by  
introducing antifields $\phi^\star $ corresponding to the all fields $\phi ( \equiv A_\mu, \bar C, C, B, \theta)$
 with opposite statistics, is given by
{\begin{eqnarray}
Z_L  = \int {\cal D}\phi\ e^{ i\int d^4x ( {\cal L}_{Th''}+{\cal L}^L_{GF+GH}[\phi, \phi^\star])}.
\end{eqnarray}}
This can, further, be written in compact form as
 \begin{equation}
Z_L = \int {\cal D}\phi\  e^{ i    W_{\Psi_L  }[\phi,\phi^\star] },
\end{equation} 
where $ W_{\Psi_L  }[\phi,\phi^\star]$ is an extended quantum action in Lorentz gauge.
The generating functional does not depend on the choice of gauge-fixing fermion \cite{ht}.
The extended quantum action for  Thirring model,
 $W_{\Psi_L  }[\phi,\phi^\star]$, satisfies  the following   mathematically rich 
 relation, called the quantum master equation \cite{wei},  
\begin{equation}
\Delta e^{iW_{\Psi_L  }[\phi, \phi^\star ]} =0  \ \mbox{ with }\ 
 \Delta\equiv \frac{\partial_r}{
\partial\phi}\frac{\partial_r}{\partial\phi^\star } (-1)^{\epsilon
+1}.
\label{mq}
\end{equation}
The antifields, which get identified  with gauge-fixing fermion in Lorentz gauge   $\Psi_L =-i\bar C\left(\partial_\mu A^\mu +\frac{\xi}{2}B\right)$, are
{ \begin{eqnarray}
A_\mu^{\star }&=&\frac{\delta\Psi_L}{\delta A^\mu}=  i\partial_\mu\bar C,
\nonumber\\
 \bar C^{ \star}&=&\frac{\delta\Psi_L}{\delta \bar C}=  -i\left(\partial_\mu A^\mu +\frac{\xi}{2}B\right),\nonumber\\ C^{\star}&=&\frac{\delta\Psi_L}{\delta C}= 0
 ,\ \ \theta^{\star}=\frac{\delta\Psi_L}{\delta \theta}= 0.
\end{eqnarray}}
Similarly, the generating functional for Thirring model in $R_\xi$ gauge  is defined, compactly, as  
{\begin{eqnarray}
Z_R & =& \int {\cal D}\phi\ e^{ i\int d^4x ( {\cal L}_{Th''}+{\cal L}^R_{GF+GH}[\phi, \phi'^\star])},\nonumber\\
&=& \int {\cal D}\phi\  e^{ i    W_{\Psi_R  }[\phi,\phi'^\star] }.
\end{eqnarray}}

The following expression for antifields, in the case of $R_\xi$  gauge, are obtained
{ \begin{eqnarray}
A_\mu^{\prime\star }&=&\frac{\delta\Psi_R }{\delta A^\mu}=   i\partial_\mu\bar C,
\nonumber\\
 \bar C'^{ \star}&=&\frac{\delta\Psi_R }{\delta \bar C}=  -i \left(\partial_\mu A^\mu+\sqrt{N}\frac{\xi}{G}\theta +\frac{\xi}{2}B\right),\nonumber\\ C'^{\star}&=&\frac{\delta\Psi_R}{\delta C}= 0,\ \ \theta'^{\star}=\frac{\delta\Psi_R}{\delta \theta}=  -i\sqrt{N}\frac{\xi}{G}\bar C,
\end{eqnarray}}
where    $\Psi_R =-i\bar C\left(\partial_\mu A^\mu+\sqrt{N}\frac{\xi}{G}\theta +\frac{\xi}{2}B\right)$ is utilized.
To connect Lorentz and $R_\xi$ gauges in BV formulation, we construct the following  infinitesimal
field-dependent parameter $\Theta' [\phi]$ 
\begin{eqnarray}
\Theta' [\phi] =i \int d^Dy \left[\bar C\bar C'^\star -\bar C\bar C^\star \right].
\end{eqnarray} 
The Jacobian of the path integral measure  in the generating functional, for   this   parameter, is computed utilizing relation (\ref{J}). 
 The resulting Jacobian factor   changes the quantum action as
\begin{eqnarray}
W_{\Psi_L }[\phi,\phi^\star]    \stackrel{FFBRST}{----\longrightarrow }
W_{\Psi_R  }[\phi,\phi^\star].
\end{eqnarray}
This reflects the validity of result at quantum levels also. Hence,
we conclude that the FFBRST transformations connect two different solutions of quantum master equation of the Thirring model.

 \section{Conclusion}
 From   the effective potential points of view, the existence of the
second order phase transition   associated with the spontaneous
breakdown of the chiral symmetry in the  $D$ ($2<D<4$) dimensional Thirring model
has been analysed. And the explicit critical number of flavors  has 
derived as a function of the four-fermion coupling constant.
The Thirring model as a gauge
theory, by introducing the Stueckelberg field as a BF field, is well studied. In this context, without gauge interactions the Thirring model is identified with the gauge-fixed version of a gauge theory and has the well known BRST symmetry even after the
gauge-fixing.

 In this paper, we have considered  
 the gauge invariant as well as renormalizable Thirring model in $D$ ($2<D<4$)  dimensions. Since, the 
 gauge invariance possesses the  unphysical degrees of freedom. According to standard 
 quantization procedure, we have to remove them by breaking the local gauge invariance.
 This can be achieved by fixing an appropriate gauge. We have discussed, the well-studied, Lorentz and $R_\xi$ 
 gauges in this context.  The remarkable properties of these gauges in  Thirring 
 model are that, in   $R_\xi$ gauge,  the   Stueckelberg (BF) field    $\theta$   is 
 completely 
decoupled to the massive vector boson independently of $\xi$ and makes the computations simple.  However, 
 in Lorentz gauge, the Stueckelberg field    $\theta$ is  decoupled to  the massive vector boson  only if $\xi=0$. In this sense, $R_\xi$ gauge is more acceptable for
 the model. 
To map these gauges, we have extended the BRST symmetry by making the transformation parameter finite
 and field dependent. Under such transformation, the path integral measure is 
 not unchanged, rather it changes in a non-trivial way.
 The Jacobian of path integral measure depends, explicitly, on field-dependent parameter.
 We compute the Jacobian for an arbitrary parameter, to be valid at a general ground.
 However, for an specific choice of parameter, we have illustrated that 
 the Jacobian connects the Lorentz gauge to $R_\xi$ gauge. Though   we have established
 a connection for a particular set of gauges, this formalism would be valid for connecting
 any two set of gauges. We have computed the extended quantum action as well as quantum master equation,
 utilizing BV formulation. Further, we have shown a mapping between the two different solutions of
 quantum master equation with the help of FFBRST transformation.
Since, the analysis of the Thirring model, as the gauged non-linear sigma model, is given  from the viewpoint of the constrained system, which implies that the present investigation might be useful from the perspectives of 
 non-linear sigma model.


\begin{thebibliography}{99}
\bibitem{a} D.K. Hong and S.H. Park, Phys. Rev.  D 49, 5507 (1994).
\bibitem{b}  T. Itoh, Y. Kim, M. Sugiura and K. Yamawaki, Prog. Theor. Phys. 93, 417
(1995).
\bibitem{c}  K.-I. Kondo, Nucl. Phys. B 450, 251 (1995).
\bibitem{wei0}S. Weinberg, Phys. Rev. D 13, 974 (1976); D 19, 1277 (1979);
L. Susskind, Phys. Rev. D 20, 2619 (1979).
\bibitem{vo} V. A. Miransky, M. Tanabashi and K. Yamawaki, Phys. Lett. B 221, 177 (1988); Mod. Phys. Lett.
A 4, 1043 (1989);
 W. J. Marciano, Phys. Rev. Lett. 62, 2793 (1989); 
W. A. Bardeen, C. T. Hill and M. Lindner, Phys. Rev. D 41, 1647 (1990).
\bibitem{3} B. Holdom, Phys. Lett. B 150, 301 (1985).
K. Y amawaki, M. Bando and K. Matumoto, Phys. Rev. Lett. 56, 1335 (1986).
T. Akiba and T. Yanagida, Phys. Lett. B 169, 432 (1986).
T. Appelquist, D. Karabali and L. C R. Wijewardhana, Phys. Rev. Lett. 57, 957 (1986).
\bibitem{4} V. A. Miransky and K. Yamawaki, Mod. Phys. Lett. A 4, 129 (1989).
K. Matumoto, Prog. Theor. Phys. 81, 277 (1989).
T. Appelquist, T. Takeuchi, M. Einhorn and L. C. R. Wijewardhana, Phys. Lett. B 220, 223 (1989).
\bibitem{u} K.-I. Kondo, M. Tanabashi and K. Yamawaki, Prog. Theor. Phys. 89, 1249 (1993);
Mod. Phys. Lett. A 8, 2859 (1993).
\bibitem{x} W. A. Bardeen, C.N. Leung and S.T. Love, Phys. Rev. Lett. 56, 1230  (1986).
\bibitem{y} C.N. Leung, S.T. Love and W.A. Bardeen, Nucl. Phys. B 273, 649 (1986).
\bibitem{z} K.-I. Kondo, H. Mino and K. Yamawaki, Phys. Rev.D 39, 2430 (1989).
\bibitem{8} Y. Kikukawa and K. Yamawaki, Phys. Lett. B 234, 497 (1989).
H.-J. He, Y.-P. Kuang, Q. Wang and Y.-I. Yi, Phys. Rev. D 45, 4610 (1992).
\bibitem{ga} D.J. Gross and A. Neveu, Phys. Rev. D 10, 3235 (1974).
 \bibitem{kond} K.-I. Kondo, Prog. Theor. Phys. 98,  211 (1997).
\bibitem{wei} S. Weinberg, {\it{ The quantum theory of fields, Vol-II: Modern
applications}} (Cambridge, UK Univ. Press, 1996).
 
 \bibitem{brst} C. Becchi, A. Rouet, R. Stora, Annals Phys. {\bf{98}} (1974) 287.
\bibitem {tyu}I. V. Tyutin, LEBEDEV-{\bf 75-39} (1975).

\bibitem{sudu} S. Upadhyay,  EPL 103, 61002 (2013); Phys. Lett. B 723, 470 (2013); Eur. Phys. J. C 74, 2737  (2014).
 \bibitem{sdj} S. D. Joglekar and B. P. Mandal, Phys. Rev. D. 51, 1919  (1995).

 
\bibitem{sdj1}  S. D. Joglekar and B. P. Mandal, Int. J. Mod. Phys. A 17, 1279 (2002).
 \bibitem{susk}   S. Upadhyay,   S. K. Rai and B. P. Mandal,  J. Math. Phys.  {52}, {022301} (2011).
  \bibitem{upad} M. Faizal, S. Upadhyay and B. P. Mandal, Eur. Phys. J. C 76, 189 (2016). 
 \bibitem{gri} V. N. Gribov,  Nucl. Phys. B 139, 1 (1978).
 \bibitem{zwan} D. Zwanziger,  Nucl. Phys. B 323, 513 (1989).
\bibitem{zwan1} D. Zwanziger, Nucl. Phys. B 399, 477 (1993).
\bibitem{sb} S. Upadhyay and B. P. Mandal,   Phys. Lett. B 744, 231 (2015);  Prog. Theor. Exp. Phys. 053B04  (2014); Eur. Phys. J.  {C 72},  2065 (2012); Annls. Phys. {  327}, 2885 (2012); Eur. Phys. Lett. {  93}, 31001 (2011);  Mod. Phys. Lett.   {A 25}, { 3347} (2010). 

 
\bibitem{smm} S. Upadhyay, M. K. Dwivedi and B. P. Mandal,  Int. J. Mod. Phys. A 30, 1550178 (2015); Int. J. Mod. Phys. A 28, 1350033 (2013).
 \bibitem{bl}
 J.F Bagger and N. Lambert, 
Phys. Rev. D 75, 045020 (2007); 
  Phys. Rev. D 77, 065008 (2008); JHEP 0802, 105 (2008).
  \bibitem{g} A. Gustavsson,  
Nucl. Phys. B 811, 66 (2009).
\bibitem{abjm}O. Aharony, O. Bergman, D. L. Jafferis and J. Maldacena,
 JHEP. 0810, 091 (2008).
\bibitem{fs} M. Faizal, B. P. Mandal and S. Upadhyay, Phys. Lett. B 721, 159 (2013).

\bibitem{sudd} S. Upadhyay and D. Das, Phys. Lett. B 733, 63 (2014).
 \bibitem{fsm}M. Faizal, S. Upadhyay and B. P. Mandal, Phys. Lett. B 738, 201 (2014);  Int. J. Mod. Phys. A 30, 1550032 (2015).

\bibitem{ssb}  B. P. Mandal,  S. K. Rai and S.  Upadhyay,
EPL { 92}, {21001} (2010).
 
 \bibitem{sud001} S. Upadhyay, Phys. Lett. B 740, 341 (2015);  Annls.   Phys. 356, 299 (2015); Mod. Phys. Lett. A 30,1550072 (2015); Annls.  Phys. 340, 110  (2014); Annls.  Phys.  344, 290 (2014); EPL 105, 21001 (2014);  EPL  104, 61001  (2013); Phys. Lett. B 727, 293 (2013).
 \bibitem{ht} M. Henneaux, C. Teitelboim, {\it{ Quantization of gauge
systems}}, Princeton, USA: Univ. Press (1992).
\bibitem{rs} R. Banerjee and S. Upadhyay, Phys. Lett. B  734, 369 (2014).



\bibitem{lav}
P.~M.~Lavrov and O.~Lechtenfeld,
 { Phys. Lett. B  725}, 382
(2013).


\bibitem{ale}
P.~Y.~Moshin and A.~A.~Reshetnyak,
 { Nucl. Phys. B   888}, 92
(2014).

\bibitem{ale1}
P.~Y.~Moshin and A.~A.~Reshetnyak,
{ Int. J. Mod. Phys. A {30}, 1550021
(2015)}.

\bibitem{mos1}
 P.~Y.~Moshin and A.~A.~Reshetnyak,
 {Phys. Lett. B { 739}, 110
(2014)}.



\bibitem{mos}
P.~Y.~Moshin and A.~A.~Reshetnyak,
  arXiv: {1506.04660}[hep-th].










\bibitem{k} B. Rosenstein, B.J. Warr and S. H. Park, Phys. Rep. 205, 59 (1991).
 \bibitem{lo} T. Fujiwara, Y. lgarashi and J. Kubo, Nucl. Phys. B 341, 695 (1990); Phys. Lett. B 261,
 427  (1990).
 \bibitem{nv} N.V. Krasnikov and A.B. Kyatkin, Mod. Phys. Lett. A 6, 1315 (1991).

\end{thebibliography}
\end{document}